\documentclass[12pt]{article}
\begin{document}
\parindent=0pt
\parskip=6pt
\rm

\begin{center}

{\Large \bf Critical behaviour of thin films\\ with quenched impurities}

\normalsize

\vspace{1cm}

{\bf L. Craco}$^{\dag}$, {\bf L. De Cesare}$^{\ddag}$, {\bf I.
Rabuffo}$^{\ddag}$, {\bf I. P. Takov}$^{\ast}$, and {\bf D. I. Uzunov}$^{\ast,
\dag, \ddag, \P}$

$^{\dag}$ Max--Planck--Institut f\"{u}r Physik Komplexer Systeme,
Aussenstelle Stuttgart, Heisenbergstrasse 1, D-70569 Stuttgart,
Federal Republic of
Germany.\\
{\em $^{\ddag}$ Dipartimento di Scienze Fisiche "E. R. Caianiello",
Universit\`{a}
di Salerno,\\ I--84081 Baronissi, Salerno, Italy, and Istituto Nazionalle per
la Fisica della Materia, Unit\`{a} di Salerno, Italy. \\
 $^{\ast}$ CPCM Laboratory, G. Nadjakov Institute of Solid State Physics,\\
Bulgarian Academy of Sciences, BG--1784 Sofia, Bulgaria.}\\
 $^{\P}$ {\em Corresponding author.}\\

\end{center}

\vspace{0.5cm}


\vspace{0.5cm}


\vspace{0.5cm}

\vspace{1cm}

{\bf Abstract}

The critical behaviour of thin films containing quenched random
 impurities
and inhomogeneities is investigated by the renormalization--group method
to the one--loop order within the framework of the $n$--component
$\phi^4$--model.The finite--size crossover in impure films has been consdered
on the basis of the fundamental relationship between the effective
dimensionality $D_{\mbox{\footnotesize eff}}$ and the characteristic lengths of
the system. The fixed points, their stability properties and the
critical  exponents are obtained and analyzed, using a $\tilde{\epsilon} = (4 -
D_{\mbox{\footnotesize eff}})$--expansion near the effective
spatial dimensionality $D_{\mbox{\footnotesize eff}}$ of the fluctuation modes
in $D$--dimensional hyperslabs with two types of quenched impurities:
point--like impurities with short--range random correlations and extended
(linear) impurities with infinite--range random correlations long the spatial
direction of the small size. The difference between the
critical properties of infinite systems and films is demonstrated and
investigated.
A new critical exponent, describing the scaling properties of the thickness of
films with extended impurities has been deduced and calculated. A
special
attention is paid to the critical behaviour of real impure films $(D=3)$.

\newpage

\normalsize

\section{Introduction}
\label{sec1}

The development of the theory of phase transitions in thin films in the
last decades
has been based mainly on
statistical
methods~\cite{Bind:Phase} such as, for example, the lattice mean--field (MF)
approximation~\cite{Pears:Rev, Uzun:Suz1}, the
phenomenological Landau--Ginzburg (LG) approach~\cite{Kag:Zh}, the general
scaling
theory~\cite{Fish:Cr, Bar:Phase}, and  renormalization--group (RG)
schemes of calculation~\cite{Bray:Phys, Uzun:Suz2}.
The general approach to the description of finite--size (FS)
systems having different geometries, including the film geometry, is the Fisher
FS  scaling
theory~\cite{Fish:Cr,Bar:Phase}. Within this framework the
field--theoretical RG  methods
are applied to the investigation of the possible types of critical
behaviour and the calculation of important characteristics
like the critical exponents, scaling amplitudes and crossover
functions~\cite{Bray:Phys, Uzun:Suz2}. The calculations are performed
for systems with a general spatial dimensionality $D$.
The predictions for real systems ($D=3$) are made by extrapolations
from the corresponding $\epsilon = (D_{U} - D)$--expansions,
 where $D_U$  is the
upper critical dimensionality~\cite{Uzun:Int}). The method of
$\epsilon$--expansions is widely and successfully applied to the
investigation of the critical properties
of FS and infinite systems. It is proven to be particularly convenient for the
study of complex models of real substances, where the competing effects play a
substantial role.

 For FS systems one of the most interesting problems is
the dimensional
FS crossover (FSC) from the usual $D$--dimensional critical behaviour to the
 corresponding
$d=(D-D_0)$--dimensional critical behaviour when the finite sizes $L_{0i}
 (i=1,...,D_0)$ are
less than the correlation length $\xi \sim (T-T_c)^{-\nu}$; $T_c$ is the
 critical temperature~\cite{Fish:Cr, Bray:Phys}.
As $\xi \to \infty$ for $T \to T_c$, the FSC $(D \to d)$ in the asymptotic
 critical behaviour
 occurs always provided the number $d$ of "infinite" dimensions $L_i \gg \xi$,
 ($i=1,...,d$) is
larger than the lower borderline (critical) dimensionality
$D_L$~\cite{Uzun:Int}.

In this paper we shall study the FSC in $D$--dimensional hyperfilms $(D_0=1,
d=D-1)$ of  thickness $L_0$
with quenched random impurities and
inhomogeneities~\cite{Uzun:Int,Grin:Fun}, which can be described by a properly
chosen $\phi^4$--model of
critical phenomena with
an $n$--component fluctuation field (order parameter) $\phi(x)=\{
\phi_\alpha
(\vec x), \alpha=1,...,n\}$.
 A special attention will be paid to the case of real
films  $(D=3)$. Thus we shall investigate a complex system with three competing
effects: the thermal fluctuations, the FS, and the disorder. The main features
of the critical behaviour in such complex systems can be described by the
Wilson--Fisher RG recursion relations in the one--loop
approximation~\cite{Uzun:Int}. The one--loop approximation leads to a reliable
prediction of the possible types of critical behaviour, namely, of the fixed
points (FP) of the RG equations, their stability properties, and of the first
corrections to the MF values of the critical exponents. We shall perform this
programme for basic models of random impurities by introducing a new approach
to the RG investigation of the FS effects.

The RG investigations of films reveal the two limiting cases of the FSC: (i)
thick (quasi--$D$--dimensional) films, where the ratio $y =
(L_{0}/\xi)$ tends to infinity $(y \gg 1)$, and (ii) thin
(quasi--$d$--dimensional) films, where the ratio $y$ tends to zero $(y \ll
1)$. In
the former case the upper borderline dimensionality is $D_{U} = 4$, whereas in
the latter case this dimensionality is $D_{U} = 5$
~\cite{Bray:Phys,Uzun:Suz2}. The $\epsilon$--expansions for the cases (i)
and (ii)
are performed for $\epsilon = (5-D) = (4-d)$ and $\epsilon = (4-D) = (3-d)$,
respectively. Here we shall present an unified RG
investigation based on an $\tilde{\epsilon} = (4 - D_{\mbox{\footnotesize
eff}})$--expansion, where $D_{\mbox{\footnotesize eff}} \in [d,D]$ is the
effective dimensionality of the fluctuation modes $\phi(\vec{x})$~\cite{DeC:J}.
For this aim we
shall apply the so--called integration over dimensionalities $\delta ( =
D_{\mbox{\footnotesize eff}} - d)$ less than unity.
 This approach has been recently used to perform a $\delta \leq 1$--integration
 over
the time axis for problems of the quantum critical behaviour in disordered
superfluids~\cite{Sch:Lett}. Working in a similar way one may obtain a
description of the
critical behaviour in impure thin films for any effective spatial
dimensionality $D_{\mbox{\footnotesize eff}} \in  [d,D]$.

The term effective
spatial dimensionality ($D_{\mbox{\footnotesize eff}}$) has been recently
introduced~\cite{DeC:J} in an attempt for a description of the FSC as a smooth
variation of
the effective dimensionality of the fluctuation modes $\phi(\vec{x})$
with the variations of the ratio $y = (L_{0}/\xi)$. Although the precise
dependence
$D_{\mbox{\footnotesize eff}}(y)$ cannot be easily obtained, it seems
intuitively obvious that in thin films $D_{\mbox{\footnotesize eff}}(y) \sim d$
for $y \ll 1$ (the case of thin films) and $D_{\mbox{\footnotesize eff}}(y)
\sim D = (d + 1)$ for $y \gg 1$ (the case of thick films or, equivalently,
of almost infinite systems).

The random impurities and inhomogeneities produce the effect of "random
critical temperature"  $T_c(\vec x)$
which depends on the spatial vector $\vec
x$ \cite{Uzun:Int,Grin:Fun,Lub:Rev,Dor:Lett}.
This type of (nonequilibrium, quenched) disorder is
investigated by choosing a convenient, usually, Gaussian distribution for the
 random
(nonequilibrium) temperature $T_c(\vec x)$. The critical
 behaviour
of infinite systems both for point~\cite{Lub:Rev} and
 extended~\cite{Dor:Lett}
impurities is well known; see also~\cite{Uzun:Int, Grin:Fun}. We consider the
FS effect on the critical behaviour in films with these basic types of
disorder.

Within the model of point impurities with spatially isotropic
short--range random correlations the FSC $(D\to d)$ can be easily proven.
Besides, we demonstrate that the critical exponents are smooth functions of
$D_{\mbox{\footnotesize eff}}$ and, hence, of the ratio $y = (L_{0}/\xi)$. We
shall show that the impure critical behaviour predicted by
predicted by T. C. Lubensky~\cite{Lub:Rev} for infinite systems with point
impurities occurs in thin
films for any effective dimensionality $D_{\mbox{\footnotesize eff}} > 2$, i.e.
for
any ratio $y = (L_{0}/\xi)$.

The disorder of type extended impurities is
described by infinitely--ranged random correlations along one or more spatial
directions. For systems of a film geometry the appropriate choice of extended
impurities is the case of one--dimensional (line) impurities orientated along
the direction of the small size $L_{0}$ and randomly distributed along the
other spatial directions. This disorder is described b a modification of the
model of short--range correlated point impurities in which the short--range
correlations along the small size are substituted with infinite--range
correlations. The length scale of the latters is much larger than both the
correlation length $\xi$ and
 the thickness $L_{0}$. So the strongly correlated along the small size
point impurities behave like continuous uniform strings. In
regards to the critical behaviour this disorder acts like point impurities with
a short--range random distribution along the large (infinite) dimensions
$L_{i}$ and an uniform distribution along the small size $L_{0}$. The
anisortopy of the random correlations lead to a quite unusual critical
behaviour of the impure films. This behaviour is described in details. As
a result of our analysis, we
have identifies a new critical exponent describing the scaling law of the
thickness $L_{0}$ of the film. This exponent is analogous to the dynamical
critical exponent $z$ in disordered classical~\cite{Grin:Maz} and quantum
systems~\cite{Kor:Uzun}.

In Section 2 we present the model of consideration. The specific FS features of
the model
are discussed in Section 3. The integration
in dimensionalities less
than unity and the effective dimensionality $D_{\mbox{\footnotesize eff}}$ are
introduced in Section 4. The RG
investigations presented in Sections 5 and 6 rely upon the basic problems
discussed in Section 3 and 4. In Section 7 we summarize
the results and discuss their applicability to real three--dimensional $(3D)$
films. The analogy between the FSC in thin films and the classical--to--quantum
crossover~\cite{Hertz:Rev} at low temperatures is also disenssed.

\section{Model}
\label{sec2}

We shall use the usual $\phi^4$--Hamiltonian $({\cal H} = H/T,
 k_B=1)$
of the $n$--component fluctuation field $\phi (\vec x)$.
We shall stick to the usual notations~\cite{Uzun:Int,Grin:Fun,Lub:Rev} in which
the Hamiltonian ${\cal
H}$ can be written in the form
\begin{equation}
\label{eq1}
{\cal H} = \frac{1}{2} \int d^D x \left \{ (\nabla \phi)^2 + r(\vec x) \phi^2
+ 2u \phi^4 \right\}\;.
\end{equation}
In Eq.~(\ref{eq1}), $u>0$, $r(\vec x)=r + \varphi (\vec x)$, where
$\varphi(\vec{x})$ is a random function intended to describe the disorder
effects and $r = \alpha_{0}(T-T_{c0})/T_{c0}$ is the usual Landau parameter
represented by
the (bare) critical temperature $T_{c0}$ of the pure systems $[\varphi(\vec x)
\equiv 0]$. The random function $\varphi(\vec{x})$ obeys
the Gaussian distribution~\cite{Uzun:Int,Grin:Fun}:
\begin{equation}
\label{eq2}
[\varphi(\vec x) \varphi(\vec x^\prime)]_R = \Delta \delta(\vec x - \vec
 x^\prime)\;,
\end{equation}
$(\Delta \geq 0)$. The function $\varphi(\vec{x})$ is related to the
"random" critical temperature $T_{c0}(\vec{x})$ in the following way:
 $T_{c0} (\vec x) = T_{c0}[1 - \varphi(\vec x)/\alpha_{0}]$. The true
(renormalized) critical temperature $T_c$  will be
a function of $T_{c0}$ and the renormalized values of the interaction parameters
 $u$ and
$\Delta$. The averages $\langle A \rangle_R$ of the physical quantities
$A[\varphi(\vec x)]$ are defined by the functional integral
\begin{equation}
\label{eq3}
\langle A \rangle_R  = \int \prod_{\vec{x}} {\cal D} \varphi (\vec x)
e^{-\frac{1}{2\Delta} \int d^D x \varphi^2 (\vec x)}
A \left[ \varphi (\vec x) \right] \;.
\end{equation}
We shall work with periodic boundary conditions.
This means that we neglect the surface energy which is important
in the investigation of other properties of thin films. This way of
treatment gives the opportunity to investigate the net effect of the FS $L_{0}$
on the critical behaviour of the impure films.

The RG investigation is
performed in the space of the wave vectors. Taking into account
the lattice structure, the components $q_{\nu}$ of the $D$--dimensional wave
vector $\vec{q} = (q_{\nu}; \nu = 0,1,...,d)$
are confined in the first Brillouin zone $(-\pi/a < q_{\nu} \leq
\pi/a)$. However, because of the long--wavelength approximation~(LWLA),
$aq_{\nu} \ll \pi$, which is unavoidably included in field models like the
Hamiltonian~(\ref{eq1}),
the upper cutoff $\Lambda$ of the wave vector $\vec{q}$ is much smaller than
$(\pi/a)$, namelly, $\Lambda = \gamma(\pi/a)$, where $\gamma \ll 1$. While
the LWLA approximation does not introduce restrictions on the dimensions
$L_{i}$ which are initially supposed to be almost infinite ($L_{i} \gg \xi$
for any $ a < \xi < \infty $), the same approximation
restricts the variations of the thickness: $a \ll L_{0}$. Therefore our
consideration is confined within the latter condition and our results could
not be extended to exactly two--dimensional ($2D$) films (mono--atomic layers).
 It is convenient to write the
$D$--dimensional wave vector $\vec{q}$ in the form
$\vec{q} = (k_{0},\vec{k})$, where the wave number $k_{0} = (2\pi n_{0}/L_{0})$
with $n_{0} = 0, \pm 1, ...$, corresponds to the thickness $L_{0}$, and the
$d$--dimensional wave vector $\vec{k}$ with components $k_{i} = (2\pi
n_{i}/L_{i})$, ($n_{i} = 0, \pm 1,...,$) corresponds to the
$d$--dimensional "subsystem" of the film.

The short--range random correlations~(\ref{eq2}) correspond to point--like
random
impurities and inhomogeneities which are equally distrubuted along all $D$
directions. In the space of the wave vectors $\vec{q}$ the Eq.~(\ref{eq2})
takes the form
\begin{equation}
\label{eq4}
[\varphi(\vec q) \varphi(\vec{q}\;']_R = \Delta \delta
(\vec{q},\vec{q}\;')\;,
 \end{equation}
where $\delta(x,y)$ is the Kronecker symbol. The disorder of type "extended
impurities"
is described by the Gaussian distribution given by
\begin{equation}
\label{eq5}
[\varphi(\vec r) \varphi(\vec{r}\;')]_R = \Delta \delta (\vec{r} -
\vec{r}\;')\;,
 \end{equation}
where the vector $\vec{r}\;$ lays in the $d$--dimensional subsystem; as given
by the definition $\vec{x}$ by $\vec{x} = (x_{0},\vec{r})$. In the $q$--space,
the Eq.~(\ref{eq5}) takes the form
\begin{equation}
\label{eq6}
[\varphi(\vec k) \varphi(\vec k^\prime)]_R = \Delta \delta
(\vec{k},\vec{k}')\;,
 \end{equation}
or, equivalently,
\begin{equation}
\label{eq7}
[\varphi(\vec q) \varphi(\vec{q}\;')]_R = \Delta \delta
(\vec{k},\vec{k}')\delta(0,\vec{k}_{0})\delta(0,\vec{k}_{0}')\;.
 \end{equation}
These models of impure systems have been briefly explained in Section 1.
Besides, they are described well in several preceding
papers~\cite{Grin:Fun,Lub:Rev} and we shall not enter in more details.

\section{Film geometry and FSC}
\label{sec3}

In order to clarify the application of the Hamiltonian~(\ref{eq1}) to FS
systems we shall consider
simple lattice sums which appear in the one--loop perturbation contributions to
the "self--energy" parameter $r$ and the interaction constant $u$:
\begin{equation}
\label{eq8}
A_{m}(r) = \frac{1}{V_d} \sum_{\vec k} S_{m}(k,r) \;,
\end{equation}
where $m = 1,2,...,$ $k =|\vec{k}|$, $V_d = (L_1,...,L_d) = V_{D}/L_{0}$, and
\begin{equation}
\label{eq9}
S_{m}(k,r) = \frac{1}{L_0} \sum_{k_0} \frac{1}{(k_0^2 + k^2 + r)^{m}} \;.
\end{equation}
In a close vicinity of the critical point $T_{c}$, where $r = \xi^{-2} \sim
0$, the upper cutoff $\Lambda$ for the wave vector $\vec{q}$ can be ignored and
the sums~(\ref{eq8}) and~(\ref{eq9}) over $\vec{k}$ and $k_{0}$ can be extended
to infinity because the essential contributions to these sums are given only by
the small wave numbers $(0 \sim q^{2} \sim \xi^{-2})$. As we are interested
in the critical behaviour in a close vicinity of the critical
point, where $\xi \gg a$, the essential contribution to the sums
$S_{m}$
and $A_{m}$ will be given by the terms with wave numbers $q_{\nu} \leq \xi$ and
therefore we must choose the cutoff $\Lambda\xi > 1$. This is consistent with
the LWLA provided $(a/\xi) <  \gamma \ll 1$. Under the latter conditions
the cutoff $\Lambda$ can be kept finite
or set infinite without any effect on the results of the summation. As we shall
apply the Wilson--Fisher RG method, it seems convenient to keep the cutoff
$\Lambda$ for the wave vectors $\vec{k}$ and neglect the upper cutoff for the
wave numbers $k_{0}$. It is important to keep in mind that our investigation of
the critical behaviour of slabs is valid only for
\begin{equation}
\label{eq10}
 \xi\Lambda \gg 1\;, \;\;\;\;\;\;\;\;  L_{0}\gg a\;.
\end{equation}

While the $\vec{k}$--summation in Eq.~(\ref{eq8}) can be always replaced by a
$\vec{k}$--integration, the
summation~(\ref{eq9}) over the wave vector component $k_{0}$ can be transformed
to
an integration only if certain conditions are satisfied.
The latter become clear from the result of the summation~(\ref{eq9}) over
$k_{0} \in (-\infty, \infty)$ for $m = 1$:
\begin{equation}
\label{eq11}
S_{1}(k,r) = \frac{L_{0}}{2y(k)} \mbox{cth} \left [\frac{y(k)}{2} \right ]\;,
\end{equation}
where $y(k) = L_{0}(k^{2} + r)^{1/2}$. For $y(k) \ll 1$, $S_{1}$
from~(\ref{eq11}) coincides with the $(k_{0} = 0)$--term in~(\ref{eq9}). In
this case, the
$D$--dimensional film has a $d = (D - 1)$ dimensional behaviour.
The quantity
$y(k)$ is much less than unity
 for any $k$ only if $(L_{0}\Lambda) \ll 1$ and $y(0) \equiv y
= (L_{0}/\xi) \ll 1$. These conditions are consistent with~(\ref{eq10})
provided $a \ll L_{0} \ll (1/\Lambda)$. The condition $(L_{0}\Lambda) \ll 1$ is
satisfied
in a broad interval of values $L_{0} > a$ because of the fact that $\Lambda \ll
(\pi/a)$. Moreover, having in mind that the small values of the wave number $k$
yield the essential contribution in the sums $A_{m}$ one may consider the
condition $L_{0} \ll \xi$ instead of $(L_{0}\Lambda) \ll 1$. This new
condition
for $y(k) \ll 1$ is weaker than $(L_{0} \Lambda) \ll 1$.
Therefore, the limiting case $y(k) \sim 0$ does not necessarily
correspond to an exactly two--dimensional ($2D$) film (a single atomic layer;
$L_{0} = a$). Rather, we have shown that the exactly $2D$ films are beyond
the scope of
the LWLA and that one of the main limiting cases of our consideration is
the quasi-- $2D$ film defined by $a \ll L_{0} \ll (1/\Lambda)$.

The second limiting case is given by $y(k) \gg 1$. The cases $y(k) \ll 1$ and
$y(k) \gg 1$ are often referred to as the limits $y(k) \to 0$ and $y(k) \to
\infty$, respecticely.
 For $y(k) \to \infty$, the Eq.~(\ref{eq11}) yields a result
that can be obtained by replacing the sum $S_{1}$ by an integral over $k_{0}$,
namely, by taking the continuum limit along the small size $L_{0}$. This
corresponds to a $D$--dimensional behaviour of the system when
all summations over the wave vector components $q_{\nu}$ can be substituted
with
an integration over the vector $\vec{q}$. The quantity $y(k)$ tends to infinity
for any $k$ only if $y \to \infty$. Thus our film is quasi--$3D$ for all $L_{0}
\gg \xi$.

The behaviour of the sum $S_{1}$ in the two
limiting cases $y(k) \to 0$ and $y(k) \to \infty$ exhibits the
dimensional FSC: $d \to D = (d + 1)$. Having in mind that
$A_{m+1} = -(\partial A_{m}/\partial r)$, $m = 1,2,...,$ we see that this
dimensional crossover is a property of all perturbation terms and, hence of the
system as a whole.

The present discussion is particularly important for RG studies of the FSC,
where the perturbation integrals are calculated at $r = 0$. In such studies the
asymptotics of the corresponding integrands are considered
and, therefore, one must be sure that the product $kL_{0}$ can
reach the limiting values $(kL_{0}) \ll 1$ and $(kL_{0}) \gg 1$ within the
limitations~(\ref{eq10})~\cite{Uzun:Suz2}. The RG integrals are given by taking
the continuum limit for the summation over $\vec{k}$ in~(\ref{eq8}) and by
confining the integration in the limits
 $k \in [\Lambda/b, \Lambda]$, where $b > 1$ is the RG rescaling
number. For such integrals with a lower cutoff $\Lambda/b$
the
condition $(\Lambda L_{0}) \gg b$ for thick (almost--$3D$) films at the
critical point $(r = 0)$
is satisfied for $L_{0} \gg (ba/\gamma\pi)$.
 The condition for thin
(quasi-- $2D$)
films is $(\Lambda L_{0}) \ll 1$ and, together with $a \ll L_{0}$, we have $a
\ll L_{0} \ll (a/\gamma\pi)$. All these conditions are
consistent within the LWLA and the requirement for a criticality as given by
inequalities~(\ref{eq10}).

Therefore the RG investigations are
consistent with both the aims and the model chosen in this paper. The behaviour
of the exactly $2D$ films is beyond the scope of our consideration but we can
reliably investigate the FSC from quasi--$3D$ to quasi--$2D$ films.

\section{Integration in noninteger dimensionalities}
\label{sec4}

An obvious disadvantage of all existing descriptions of the
FSC and other dimensional crossovers~\cite{Hertz:Rev} is that the limiting
cases can be
easily proven and described but the intermediate case ($ 0 < y < 1$) presents a
difficult and unresolved task. The
systematic
way of investigation of the intermediate cases ($y \sim 1$) is to use the
Euler--Maclaurin
summation formula and take into account the corrections in inverse powers of
$y(k)$ to the continuum limit. Such a treatment requires a
numerical analysis. Alternatively, one may perform the RG studies by an
integration in noninteger dimensionalities. In this Section we shall discuss
the advantages and the disadvantages of this method. We shall show that it can
be applied as an interpolation between the limiting cases $y \ll 1$ and $y \gg
1$ only to specific theoretical schemes such as the RG.

\subsection{$\delta$--integration}
\label{sec4.1}
Let us consider the integral
\begin{equation}
\label{eq12}
A_{1}(r,b) = L_{0} \int
\frac{d^{d}k}{(2\pi)^{d}}\frac{\mbox{cth}[y(k)/2]}{2y(k)}\;,
 \end{equation}
which follows from~(\ref{eq8})
and~(\ref{eq11}). The integrand $S_{1}(k,r)$
exhibits a single--power behaviour
$[\sim y^{\sigma}(k)]$ only for $y(k)
\ll 1$
and $y(k) \gg 1$. The existence of a leading power dependence of the integrals
$A_{m + 1} = -(\partial A_{m}/\partial r)$ on $y = (L_{0}\sqrt{r})$ and the
irrelevanve of the correction terms lead to a simple structure of the RG
equations and, hence, to their scale invariant solutions, which are important
for the description of the critical behaviour. The problem is to achieve such
solutions for the intermediate cases of $y \sim 1$, too.

It is impossible to construct an exact integral counterpart of the
Eq.~(\ref{eq12}) with a power
law behaviour with respect to $y$ and for this reason, here we shall consider
an approximate solution of the problem. We shall substitute the sum
$S_{1}(k,r)$ with the $\delta$--dimensional integral
\begin{equation}
\label{eq13}
S'_{1} = L_{0}^{\delta - 1} \int
\frac{d^{\delta}x}{(2\pi)^{\delta}}\frac{1}{(x^{2} + k^{2} + r)}\;.
\end{equation}
Accordingly, the integral~(\ref{eq12}) will be substituted by the double
$(\delta,d)$--dimensional integral
 \begin{equation}
\label{eq14}
A'_{1}(r,b) =  L_{0}^{\delta - 1}
 \int \frac{d^{d}k}{(2\pi)^{d}}\int
\frac{d^{\delta}x}{(2\pi)^{\delta}}
\frac{1}{(x^{2} + k^{2} +
r)}\;.
 \end{equation}
The alternative is to substitute the integral~(\ref{eq12}) with the $(d +
\delta)$--dimensional integral
\begin{equation}
\label{eq15}
A''_{1}(r,b) =  L_{0}^{\delta - 1} \int
\frac{d^{d + \delta}q}{(2\pi)^{d + \delta}}\frac{1}{(q^{2} + r)}\;,
 \end{equation}
where $\delta \in [0,1]$. Certainly, these substitutions are not exact
counterparts of the original quantities and their utility in our attempts to
present a reliable interpolation between the limiting cases should be
justified.

The integral~(\ref{eq15}) can be deduced after the
conjecture
that the wave vector component $k_{0}$ is a $\delta(<1)$--dimensional
(sub)vector, $\vec{k}_{0} = (k_{\mu 0}$;
 $\mu =
1,...,\delta)$ and, accordingly, that the total
wave vector $\vec{q} =\{k_{\mu 0};k_{i}\} $ and the volume
$V_{D}= L^{\delta}V_{d}$ in the
sum~(\ref{eq8})
correspond to a  $(d + \delta)$--dimensional
system~\cite{DeC:J,Sch:Lett,Dor:Lett}.
The integrals $A''_{1}$ and $A''_{(m + 1)} = -(\partial A''_{m}/\partial r)$
 have
been used in preceding studies of quantum systems~\cite{Sch:Lett} and extended
impurities~\cite{Dor:Lett} for both $\delta > 1$ and $\delta < 1$.

The
integrals $S'_{1}$ and $A'_{(m + 1)} = - (\partial A'_{m}/\partial r)$ given
by~(\ref{eq13}) and~(\ref{eq14}) are
defined with the help of another conjecture, namely, that one may perform a
smooth interpolation between the integral values in the continuum limits
$(\delta = 0)$ and $\delta = 1$ for
the $d$-- and $D = (d + 1)$--dimensional cases, respectively, with the help of
the formal rule
\begin{equation}
\label{eq16}
\frac{1}{L^{\delta}_{0}}\sum_{k_{0}} \to \int
\frac{d^{\delta}x}{(2\pi)^{\delta}} \equiv
K_{\delta}\int^{\infty}_{0} dx.x^{\delta - 1} \; ,
\end{equation}
where $K_{\delta} = 2^{1 -
\delta}/\pi^{\delta/2}\Gamma(\delta/2)$.
 Using~(\ref{eq8}),~(\ref{eq9})
and~(\ref{eq16}) one immediately
obtains the integrals~(\ref{eq13}) and~(\ref{eq14}).
The limit $\delta \to 0$
in the last integral (in spherical
coordinates) in~(\ref{eq16}) should be taken with a special attention because
of the divergency
of the gamma function $\Gamma(\delta/2)$. At first one should perform the
integration over $x$ of the integrand, say, $x^{+0}f(x)/x$, and then to take
the
limit $\delta \to 0$. Usually, the integrands [$\sim f(x)$] which appear by
the perturbation series are such that no divergences arise in the final
results for $\delta \sim 0$. This is confirmed by a direct calculation of the
integrals~(\ref{eq13}),~(\ref{eq14}) and $A'_{m+1} = - (\partial
A'_{m}/\partial r)$.

 For $\delta \to 0$ and
$\delta \to 1$, the integrals~(\ref{eq13}) and~(\ref{eq14}) exactly reproduce
the results from~(\ref{eq11}) and~(\ref{eq12}) for $y \to 0$ and $y \to
\infty$,
respectively. The same is valid for the integral~(\ref{eq15}) with respect to
Eq.~(\ref{eq12}). So there are some grounds for the supposition that the
intermediate states ($y \sim 1 $) could be interpolated by the values $0 <
\delta < 1$.

\subsection{Effective dimensionality of the fluctuation modes}
\label{sec4.2}

The coincidence of the results for the original
integrals $A_{m}(r)$ in the limits $y \to 0$ and $y \to \infty$ with the
results from the integrals $A'_{m}$ and $A''_{m}$ could be used as a
basis of the supposition that there exists a continuous increasing function
$\delta(y)$ with the properties $\delta(y\to 0) \to 0$ and $\delta(y\to \infty)
\to 1$. This supposition presents the opportunity
to intruduce a new dimensionality -- the spatial dimensionality of the
fluctuation modes $\phi(\vec{q})$ given by
\begin{equation}
\label{eq17}
 D_{\mbox{\footnotesize eff}} (y) = d + \delta (y)\; .
\end{equation}
The Eq.~(\ref{eq17}) is a straightforward generalization of the known (from
previous FS studies~\cite{Bray:Phys,Uzun:Suz2}) fact that the FS system
abruptly
changes its $D$--dimensional behaviour to the corresponding $d$--dimensional
behaviour when the thickness $L_{0}$ is lowered to values less than $\xi$.
According to the model~(\ref{eq1}), the system is represented by the field
$\phi(\vec{x})$. In
this respect the notion for the effective length $D_{\mbox{\footnotesize eff}}$
is not new but in this Section we consider it more explicitly within the
generalized form given by Eq.~(\ref{eq17}). Besides, in Section 5 and 6 we
shall introduce an  $\tilde{\epsilon} =
(D_{\mbox{\footnotesize eff}} - D^{(U)}_{\mbox{\footnotesize eff}})$--expansion
around the upper borderline effective dimensionality
$D^{(U)}_{\mbox{\footnotesize eff}}$. In general, this is a way to describe the
RG scaling in terms of the fundamental ratio $y = (L_{0}/\xi)$. Unfortunately,
the present investigation does not give an opportunity to obtain the function
$\delta(y)$ and, hence, $D_{\mbox{\footnotesize eff}}(y)$. The reason is in the
pecuriality of the approach based on the $\delta$--integrations introduced in
Section 4.1.

\subsection{Validity}
\label{sec4.3}

 Although the formal difference in their
definitions,
the integrals $A'_{m}(r)$ and $A''_{m}$ lead to the same results in many
practical calculations. An example of such a calculation will be presented in
this Section; for other examples, see Sections 5 and 6. The problem
is that the values of the
original sum~(\ref{eq11}) and integral~(\ref{eq12}) do not coincide with
the values of the corresponding integrals
in noninteger dimensinalities ($\delta \neq 0$). Therefore, the application of
the integration in noninteger dimensionalities to concrete problems requires a
special attention. The relaibility of such applications should be justified for
any particular case.  Despite the numerous RG
studies which have been already performed with the help of the integrals
$A''_{m}$, the question about the limitations of the corresponding results has
not been considered.
In order to justify our RG investigations in Sections 5 and 6,
we shall consider this problem. Besides, we shall demonstrate the degree
of the approximation by compating the original sum $S_{1}(k,y)$ and integral
$A_{1}(y)$, denoting them as functions of $y$, with the corresponding integrals
$S'_{1}(k,y)$, $A'_{1}(y)$, and $A''_{1}(y)$.

By calculating the
integral~(\ref{eq13}) for $S'_{1}$ and by comparing the result with $S_{1}$
from~(\ref{eq11}) we obtain
\begin{equation}
\label{eq18}
\mbox{th}[y(k)/2] \leftrightarrow g_{\delta}[y(k)]\;
\end{equation}
with
\begin{equation}
\label{eq19}
g_{\delta}[y(k)] = A(\delta) \left [\frac{y(k)}{2} \right
]^{1-\delta}\;,
\end{equation}
where $A(\delta) = \pi^{\delta/2}/\Gamma(1 -\delta/2)$; hereafter the
symbol "$\leftrightarrow$" will denote a "comparison" and nothing else. The
comparison~(\ref{eq18}) can be done for any $k < \Lambda$. For
$ k = 0$, we have $y(0) = y$, for $k\xi = 1$, $y(1/\xi) = \sqrt{2}$, and for a
third
wave number $k\xi=\sqrt{3}$, which also has an essential contribution in the
integrals over $\vec{k}$, we have $y(\sqrt{3}/\xi) = 1$. As we shall see
our approximations are not precise enough in order to distinguish between the
fits of $S_{1}$ and $S'_{1}$ for different wave numbers. Besides, there is a
strong argument that the most important value of $k$ at which we should make
the comparison~(\ref{eq18}) is $k = 0$. This value corresponds to the uniform
mode $\phi(\vec{k} = 0)$ describing the spontaneous symmetry breaking in the
$d$--dimensional subsystem.

 By setting $k = 0$ in~(\ref{eq18}) and~(\ref{eq19}) we have
\begin{equation}
\label{eq20}
\mbox{th}(y/2) \leftrightarrow  A(\delta) \left (\frac{y}{2} \right
)^{1-\delta}\; .
\end{equation}
The l.h.s and the r.h.s. of~(\ref{eq20}) are depicted in Fig. 1. From one side
it is obvious that $S'_{1}(0,y)$ is a good approximation to $S_{1}(0,y)$
in quite broad intervals of values of $y$, for example: $ 0 <
y(k) < 1$ and $y(k) > 4$. From the other side, it becomes evident that the
values of $\delta$ that give the best fit of $S'(0,y)$ to $S(0,y)$
are: $\delta
= 0$ -- for the case $y < 1$, and $\delta = 1$ -- for the case $y \gg 1$. This
means that, within the present consideration, the function $\delta(y)$ can be
approximated with zero for all $y < 1$ corresponding to a relatively good fit
of the curves
$\mbox{th}(y/2)$ and $g_{\delta}(y)$ and that, owing to the same arguments,
$\delta(y) \approx 1$ for $y \gg 2$. In a broad region of values $y$ around $y
= 2$ the approximation of $S(0,y) \sim S'(0,y)$ breaks down and we
 could not deduce any reliable conclision about the exact critical value
$y_{c}$ of
$y$ at which the FSC occurs. According to the present picture one may speculate
that this value is probably $y_{c} = 2$ whereas the intuitively appealing value
is $y = 1$. The value $y_{c} = 2$ comes from the factor $(1/2)$ in the front of
$y(k)$ in Eqs.~(\ref{eq18}),~(\ref{eq19}), and~(\ref{eq20}). The picture
outlined from the comparison~(\ref{eq20}) is valid for the
comparison denoted in~(\ref{eq19}). In this case one should change $y$ with
$y(k)$. For $k = (\sqrt{3}/\xi)$, $y_{c} = 1$, and the points of intersection
of $g_{\delta}$--lines will be located around the coordinate $y_{c} = 1$.

The calculculations within the present approach yield that the
locations of the minimal values of the difference between the r.h.s. and the
l.h.s. of~(\ref{eq20}) are $\delta = 0$ for all $y < 2$ and $\delta = 1$ for
all $y > 1$, namely, that $\delta(y)$ coincides with the $\Theta$--function
$\Theta(y-2)$. Certainly this qualitatively wrong result is due
to the obvious fact, see Fig. 1, that in a broad region around the point $y_{c}
= 2$, the approximation of $S_{1}(0,y)$ to $S'_{1}(0,y)$ is not valid. It can
be however concluded from such rough considerations that the real function
$\delta(y)$ will have a steep increase from very small values $\delta \sim 0$
up to
$\delta \sim 1$ in a relatively close vicinity
($|y-y_{c}| \sim y_{c}$) of the real critical point $y_{c}$.

The comparison of the integral $A_{1}(r)$
with $A'_{1}(r)$ and $A''_{1}(r)$ is more difficult. In order
to avoid
the cutoff $(\Lambda$--) dependence of the results and simultaneously to avoid
irrelevant to our problem ultraviolet divergences we shall set $\Lambda =
\infty$ and consider the differences of type $\Delta
A_{1}(y) = [A_{1}(0) - A_{1}(y)]$ rather than the integrals themselves.
Using
Eq.~(\ref{eq12}) we obtain
\begin{equation}
\label{eq21}
\Delta A_{1}(y) = \frac{1}{2}K_{d}L^{1 - d}_{0}I_{d}(y)\;
\end{equation}
with
\begin{equation}
\label{eq22}
I_{d}(y) = \int^{\infty}_{0}dz\:z^{d-1}\left [\frac{\mbox{cth}(z/2)}{z} -
\frac{\mbox{cth}(\sqrt{z^{2} + y^{2}}/2)}{\sqrt{z^{2} + y^{2}}} \right ]\;,
\end{equation} %
where $z = L_{0}k$. The difference $\Delta A'_{1}(y) = [A'_{1}(0) - A'_{1}(y)]$
is obtained from Eqs.~(\ref{eq8}) and~(\ref{eq14}):
\begin{equation}
\label{eq23}
\Delta A'_{1}(y) = A(\delta) K_{d}L^{1
- d}_{0}J'_{d}(\delta, y)\;,
\end{equation}
where
\begin{equation}
\label{eq24}
J'_{d}(\delta, y) = \int^{\infty}_{0}dz\;z^{d-1} \left [\frac{1}{z^{2 -
\delta}} - \frac{1}{(z^{2} + y^{2})^{1 - \delta/2}} \right ]\;.
\end{equation}
Finally, from Eqs.~(\ref{eq8}) and~(\ref{eq15}) we have that the difference
$[A''_{1}(0) - A''_{1}(y)]$ is given by
 \begin{equation}
\label{eq25}
\Delta A''_{1}(y) = K_{d + \delta}L^{1- d}_{0}y^{2}
\int^{\infty}_{0}dz\:\frac{z^{d + \delta -3}}{z^{2} + y^{2}}\;.
\end{equation}
The integrals~(\ref{eq22}),~(\ref{eq24}) and~(\ref{eq25}) have obvious
ultraviolet and infrared divergences at the corresponding lower and upper
borderline dimensionalities.
In order to avoid unnecessary complications in our calculation we
 shall consider the case of real films $(d = 2)$.

For $(d = 2)$ the Eqs.~(\ref{eq25}) becomes
\begin{equation}
\label{eq26}
\Delta A''_{1}(y) = \frac{A(\delta)}{2\pi L_{0}}
\left (\frac{y^{\delta}}{\delta} \right )\;.
\end{equation}
Because of the obvious infrared divergence at $\delta = 0$, we shall consider
the derivative $(\partial \Delta A''_{1}/\partial y)$ instead of the difference
$ \Delta A''_{1}$ itself:
\begin{equation}
\label{eq27}
\frac{\partial \Delta A''_{1}(y)}{\partial y} = \frac{A(\delta)y^{\delta
- 1}}{2\pi L_{0}}\;.
\end{equation}
The ultraviolet divergences in the two parts of the integral $J'_{2}(\delta,y)$
are exactly compensated each other and this integral takes the simple form
$J'_{2}(\delta,y) = y^{\delta}/\delta$. Thus we obtain that the derivative of
$\Delta A'_{1}(y)$ is equal to the derivative~(\ref{eq27}).
The integral $\partial I_{2}(y)/\partial y$ can be represented in
the form
\begin{equation}
\label{eq28}
\frac{\partial I_{2}(y)}{\partial y} = y \int_{y}^{\infty}dt \left
(\frac{\mbox{cth}(t/2)}{t^{2}} + \frac{1}{2t\mbox{sh}^{2}(t/2)} \right )
\end{equation}
which directly yields the result $\mbox{cth}(y/2)$.
For $d = 2$, the derivative of the difference $\Delta A_{1}(y)$ becomes
\begin{equation}
\label{eq29}
\frac{\partial A_{1}(y)}{\partial y} = \frac{1}{4\pi L_{0}}\mbox{cth} \left
( \frac{y}{2} \right )\;.
 \end{equation}
The correspondence~(\ref{eq20}) straingtforwardly follows from the comparison
of the Eqs.~(\ref{eq27}) and~(\ref{eq29}).

Thus we have shown that
the derivatives of the integrals $A'_{m}$
and $A''_{m}$ are quite different the derivatives of the original
integrals $A_{m}$, in particular for $y \sim y_{c}$. Having in mind that the
values of all integrals coincide
at the limiting points $\delta(y = 0) = 0$ and $\delta(y = \infty) = 1$, the
same conclusion is true for the integrals themselves.

The demonstrated deviation of the integrals in noninteger dimensionalities from
the initial integrals $A_{m}$ does not mean
that the RG analysis based on them is unreliable for $0 < \delta < 1$. The
argument here is that the RG predictions about the critical behaviour
follow from the RG transformation which reflects the structure and the symmetry
of the
Hamiltonian rather than from the values of the perturbation integrals. The
latters
determine the location of the fixed points (FPs) of the RG equations and,
therefore, might be of interest only in problems of special interest as to the
question of whether a particular stable FP is accessible by the RG flows or
not. For such special questions which often arise in investigations of complex
systems the wrong determination of the FP coordinates, given by the integrals
$A'_{m}$
or $A''_{m}$, may produce wrong conclusions. Hopefully, our RG analysis in the
next two Sections do not come upon such problems and we can arrive at reliable
predictions about the critical behaviour of impure films irrespectively on the
incorrest predictions about the location of the FPs.

In Sections 5 and 6 we
shall perform the RG investigation with the help of the integrals $A'_{m}$.
This more difficult variant is chosen to demonstrate that the results do not
depend on the particular scheme of calculation.

\section{RG analysis: short--range impurity correlations}
\label{sec5}

The conventional RG treatment of the FSC has been presented in details in
preceding works~\cite{Bray:Phys,Uzun:Suz2}, where the perturbation integrals
 $A_{m}$ have been calculated by a direct summations over $k_{0}$; see,
e.g. Eqs.~(\ref{eq11}) and (\ref{eq12}).
 We shall show that the application of
the integration in noninteger dimensionalities yields more information about
the critical properties of films than the standard RG treatment of FS
systems. For this aim we shall use the double $(\delta,d)$--integration.

The RG recursion relations can be derived in two equivalent ways: by
introducing an
initial rescaling of the wave vector component $k_{0}$ and without
such an initial rescaling~\cite{Bray:Phys, Uzun:Suz2}. We choose the
latter variant in which the rescaling
of the wave number $k_{0}$ or, equivalently, of the thickness $L_{0} \sim
(1/k_{0})$, will appear as a result of the RG transformation.
 The initial rescaling transformations are $k_{i}' = bk_{i}$ and
\begin{equation}
\label{eq30}
\phi_{\alpha}(k_{0}, b^{-1}\vec{k}) = b^{1 -
\eta/2}\phi_{\alpha}'(k_{0},\vec{k})\;,
 \end{equation}
where $b > 1$ is the rescaling number.
In our investigation within the one--loop approximation the Fisher exponent
$\eta$ is equal to zero. The calculations are carried out by the direct
way~\cite{Uzun:Int,Lub:Rev}
 of the  averaging of the random
functions.

 Using the standard way of calculations we have derived the
RG recursion relations for short--range correlated point impurities
~(\ref{eq4}) in the form \begin{equation}
\label{eq31}
L_{0}' = b^{-1}L_{0}\;,
 \end{equation}
\begin{equation}
\label{eq32}
r' = b^{2} \{ r + [4(n+2)u - \Delta]A'_{1}(r,b)\}\;,
 \end{equation}
\begin{equation}
\label{eq33}
u' = b^{3 - d} \{ u - [4(n+8)u^{2} - 6u\Delta]A'_{2}(0,b)\}\;,
 \end{equation}
\begin{equation}
\label{eq34}
\Delta' = b^{3 - d} \{ \Delta - [8(n+2)u\Delta -
4\Delta^{2}]A'_{2}(0,b)\}\;,
\end{equation}
where the integrals $A'_{m}(r,b)$ are the continuum limits of the
sums~(\ref{eq8}) with the upper cutoff $\Lambda = 1$ and an lower cutoff
$b^{-1}$; hereafter we shall omit the $(')$ of these integrals. The calculation
of the integrals to
first order in the small parameter $r$
  at the upper borderline
dimensionality $[D^{(U)}_{\mbox{\footnotesize eff}} = (d + \delta)_{U} = 4]$
yields: \begin{equation}
\label{eq35}
A_{1}(r,b) = A_{1}(0,b) - A_{2}(0.b)r\;
\end{equation}
with
\begin{equation}
\label{eq36}
A_{1}(0,b) = L_{0}^{\delta - 1}\tau(\delta) \left ( \frac{1 - b^{-2}}{2 -
\delta} \right )\;
\end{equation}
and
\begin{equation}
\label{eq37}
A_{2}(0,b) = L_{0}^{\delta - 1} \tau(\delta)\mbox{ln}b\;,
\end{equation}
where
\begin{equation}
\label{eq38}
\tau(\delta) = (1 - \frac{\delta}{2})2A(\delta)K_{4 - \delta}\;.
\end{equation} %
Further, we substitute the results~(\ref{eq35})~--~(\ref{eq37}) in
the relations~(\ref{eq32})~--~(\ref{eq34}) and perform the
change of the parameters $u$
and $\Delta$ with
$\tilde{u}_{0} = L_{0}^{\delta - 1}\tau(\delta)u_{0}$ and $\tilde{\Delta} =
L_{0}^{\delta - 1}\tau(\delta)\Delta$. In the new notations, the
Eqs.~(\ref{eq32})~--~(\ref{eq34}) take the form
\begin{equation}
\label{eq39}
r' = b^{2} \{ r +  [4(n+2)\tilde{u}- \tilde{\Delta}]\left
(\frac{1 - b^{-2}}{2 - \delta} - r\mbox{ln}b\right )\;,
\end{equation}
\begin{equation}
\label{eq40}
\tilde{u}' = b^{4 - d - \delta} \{ \tilde{u} - [4(n+8)\tilde{u}^{2}
- 6\tilde{u}\tilde{\Delta}]\mbox{ln}b\}\;,
\end{equation}
\begin{equation}
\label{eq41}
\tilde{\Delta}' = b^{4 - d - \delta} \{ \tilde{\Delta} -
[8(n+2)\tilde{u}\tilde{\Delta} - 4\tilde{\Delta}^{2}]\mbox{ln}b\}\;.
\end{equation}
The relation~(\ref{eq31}) describes the trivial scale invariance of the
thickness $L_{0}$. This triviality is a result of the equivalence of the
short--range random correlations along all $D$ spatial directions (isotropy
of the disorder). As we shall see in Section 6, the extended impurities break
this isotropy and the RG relation for $L_{0}$ becomes nontrivial. The
structure of Eqs.~(\ref{eq39})~--~(\ref{eq41})
is the same as that of the known RG equations for infinite
systems~\cite{Lub:Rev}. The difference is in the parameter
$\delta \neq 0$. We shall see in the next Section that this simple
structure of the RG equations reflects the above mentioned isotropy in
the distribution of the disorder. Note, that the factor $1/(2 - \delta)$ in
Eq.~(\ref{eq39})
is relevant only for the precise determination of the FP value of the parameter
$r$.The same iz valid for the $\delta$--dependence of the
integrals~(\ref{eq35})~--~(\ref{eq37}).
In Section 4.3 we have already mentioned that the critical behaviour
represented by the critical exponents does not depend on the concrete values of
the
FP coordinates in the parameter space $(r,u,\Delta)$ or, alternatively, in $(r,
\tilde{u},\tilde{\Delta})$.

The $(d,\delta)$--integration is a
relatively clumsy variant of calculation and cannot be easily applied in RG
calculations in higher loop
approximations. But the same results are obtained by the ($d +
\delta$)--integration~(\ref{eq15}) which
can be straightforwardly extended to considerations in higer orders of the loop
expansion. Thus one can directly obtain the two--loop results for the same
problem. The structure of the RG equations is again the same as that for the
impurity problem in the corresponding infinite systems~\cite{Lub:Rev}. The
simple
 difference is that one should change $d$ with $(d + \delta)$, as it is
for the Eqs.~(\ref{eq39})~--~(\ref{eq41}).

Thus we arrive at the main
conclusion that irrespectively to the considered order of the loop
expansion, the
results for the impure films are obtained from those for infinite systems by
the simple substitution of $\epsilon = (4 - d)$ with $\tilde{\epsilon} = [4 -
(d + \delta)] = (4 - D_{\mbox{\footnotesize eff}})$. The same is valid for the
FP coordinates. Alternatively,
by using the $\tilde{\epsilon}$--expansion one obtains the results for both the
$D$--dimensional hyperslab $(\delta > 0)$  and the corresponding infinite
system $(\delta = 0)$. For the case of slabs we use the
$\tilde{\epsilon} = (d + \delta)$--expansion as shown in
Eqs.~(\ref{eq39})~--~(\ref{eq41}) and for the infinite system we take the limit
$\delta \to 0$, which means to perform an expansion in $\tilde{\epsilon}
\equiv \epsilon = (4 - d) = (5 - D)$. The reason for this $(5-D)$--expansion
for
the infinite system is in the FSC $(d \to D)$ denoting the equivalence between
the critical behaviour in $D$--impure films and that in $d = (D -
1)$--dimensional impure infinite systems. Alternatively, in case of infinite
systems, one may take the limit $\delta \to 1$, $D_{\mbox{\footnotesize eff}} =
(d + \delta) \to D$ and, therefore, the expansion parameter is $\epsilon = 4 -
D$.

 Within the present RG analysis we
have shown that the FSC can be described as a smooth change of the behaviour of
the system with the variation of $\delta$ from $\delta = 0$ to $\delta \to 1$.
Bearing in mind our discussion in Section 4, the variations of $\delta$ between
these limiting values mean a corresponding variation of the ratio $y =
(L_{0}/\xi)$ from $y \ll 1$ to $y \gg 1$. Although the smooth variation of
$\delta$ in the RG equations in the broad interval $(0,1)$, the FS crossover is
expected to occur mainly in the interval $|y - y_{c}| \sim
y_{c}$, where the $y_{c} \sim 1$ is the critical ratio.

In general, the critical behaviour of
the hyperslab  will correspond to the MF description for geometrical
dimensionalities $ D > 5 - \delta$. As our RG results are valid above the lower
effective borderline dimensionality $(d + \delta)_{L} = 2$, the nontrivial
 impure (for $n < 4$) or pure (for $ n > 4$) stable critical behaviour will
occur for geometrical
dimensionalities $(3 - \delta) < D < (5 - \delta)$. We shall not derive and
discuss the FPs and the related critical and stability exponents of the impure
films because we have demonstrated that these quantities and, therefore, the
critical behaviour of the film as a whole, are obtained by setting $\epsilon =
(4 - d) \to \tilde{\epsilon} = (\epsilon - \delta)$ in the known results
for the respective quantities of impure infinite systems with the same type of
disorder~\cite{Lub:Rev}.

For real slabs
($D = 3)$ we find that the nontrivial critical behaviour will appear for
$\delta > 0$. This implies $L_{0} \gg a$ (quasi--$2D$ films) and, moreover,
the real
films should be thick enough to ensure a parameter $\delta > 0$. In view of
the discussion in Section 4, this probably corresponds to the ratio
$(L_{0}/\xi)$ of
order of the unity. Denoting $\tilde{\epsilon} = (\epsilon - \delta)$, where
$\epsilon = (4 - d)$, we see that the results initially obtained for
$\tilde{\epsilon} \ll 1$ can be extrapolated to real films $(d = 2)$ provided
we extend the values of our small parameter to $ 1 < \tilde{\epsilon} = (2 -
\delta) < 2$. This is an advantage with respect to the simple extension of
$\epsilon = (4 - d)$ to $\epsilon = 2$ for $2D$--films. There are no problems
for the extrapolations of the $\tilde{\epsilon}$--results because the RG
analysis does not demonstrate any peculiarities of the RG analysis at
these relatively low dimensionalities such as: the appearance of new FPs,
any runway of the knoun (at $\tilde{\epsilon} \ll 1$) FPs, a qualitative
change
of the stability properties of the FPs or, a change of the location of the FPs
in domains of the parameter space, where they could be unaccessible for the
RG flows.

From one side, we have shown, that the scope of
the RG investigations with the help of field theretical models does not include
$2D$--dimensional films (mono--layers) or extremely thin films consisting of
several atomic layers $(L_{0} \sim a)$. From the other side, it has been
completely clarified that the present approach can be undoubtedly applied to
real films $(D = 3)$.

\section{RG for extended impurities}
\label{sec6}

\subsection{RG equations}
\label{sec6.1}

The RG equations for the case the extended impurities and inhomogeneities
described by the Eqs.~(\ref{eq5})--(\ref{eq7}) are
derived in the way outlined in the preceding Section. In order to simplify the
notations we introduce the new variables $v = L_{0}^{\delta -
1}\tau(\delta)u$ and $\mu = K_{4 - \delta}\Delta$. Then, the RG relations
can be written in the form
\begin{equation}
\label{eq42}
k_0^\prime = b k_0 \left\{ 1 + \mu \frac{b^\delta -1}{2\delta} \right\} \;,
\end{equation}
\begin{eqnarray}
\label{eq43}
r^\prime & = &  b^2 \left\{ r + 4(n+2)v\left[ \frac{1-b^{-2}}{2-\delta} -r \ln b
 \right]
\right. \nonumber \\
& - & \left.  \mu \left[ \frac{1-b^{-2+\delta}}{2-\delta} - r \left(
 \frac{b^\delta -1}{\delta}
\right) \right] \right\} \;,
\end{eqnarray}
\begin{equation}
\label{eq44}
v^\prime=b^{4-d} \left( \frac{L_0^\prime}{L_0} \right)^\delta \left\{ v - 4(n+8)
 v^2 \ln b
+ 6v \mu \left( \frac{b^\delta -1}{\delta} \right) \right\} \;,
\end{equation}
\begin{equation}
\label{eq45}
\mu^\prime = b^{4-d} \left\{\mu - 8(n+2) v \mu \ln b + 4\mu^2 \left(
 \frac{b^\delta -1}
{\delta} \right) \right\} \;.
\end{equation}
These RG relations have been obtained by the $\epsilon = (4 -
D_{\mbox{\footnotesize eff}})$--expansion around the upper borderline
dimensionality $D^{(U)}_{\mbox{\footnotesize eff}} = 4$ (see Section
6). In contrast to the case in Section 6 the scaling invariant solutions
of these RG relations cannot be found for any $\delta \in [0,1]$. In order to
obtain such solutions we must consider $\delta \ll 1$ and to substitute the
factor $(b^{\delta} - 1)/\delta$ with $\mbox{ln}b$.
As our expansion parameter is $\tilde{\epsilon} = [4 - (d + \delta)] \ll 1$,
this means that we must investigate the case when $\delta \ll $ and $\epsilon =
(4 - d) \ll 1$. Thus we have a single $\tilde{\epsilon}$--expansion with two
small parameters: $\epsilon = (4 - d)$ and $\delta$. This variant of the theory
is often reffered to as a "double $(\delta,
\epsilon)$--expansion"~\cite{Grin:Fun, Dor:Lett}. The anisotropy of the random
correlations "breaks" the single $\tilde{\epsilon}$--expansion to a double one.

The $\mu$--contribution in the RG relation~(\ref{eq30}) for $k_{0}$ comes from
a perturbation term shown diagrammatically in Fig. 2. In contast to the usual
one--loop results, this self--energy contribution depends on the wave
number $k_{0}$. The reason is that the broken line of the diagram in Fig. 3
does not carry external wave numbers $k_{0}$. This is a direct consequence of
the fact that the random function $\varphi(\vec{k})$ does not depend on
$k_{0}$, i. e. of the infinite ranged correlations along the small site
$L_{0}$. The $\mu$--dependence in the RG equation
for $k_{0}$ leads to a nontrivial scaling relation for the thickness $L_{0}$,
namely, \begin{equation}
\label{eq46}
L_0 = L_0^\prime b^{1 + \mu/2} \;,
\end{equation}
 which is quite similar to the
scaling relations for the temperature $T$ known from studies of quantum
critical phenomena~\cite{Hertz:Rev}. The scaling law~(\ref{eq34}) has the
critical exponent
\begin{equation}
\label{eq47}
z_d = 1 + \frac{\mu^\ast}{2} \;,
\end{equation}
where the asterisk $(\ast)$ denotes any FP value of $\mu$. This
exponent is
analogous to the dynamical critical exponent
in the
theory of dynamical critical phenomena in classical models with quenched
disorder~\cite{Grin:Maz} and quantum systems~\cite{Kor:Uzun,Hertz:Rev}.

Using the relation~(\ref{eq46}) and
 $\delta \ll 1$,
the Eqs.~(\ref{eq43})--(\ref{eq45}) can be written in the form
\begin{equation}
\label{eq48}
r^\prime = b^2 \left\{ r + \left[ 4(n+2)v - \mu \right] \left[
 \frac{1-b^{-2}}{2} - r\ln b
\right] \right\} \;,
\end{equation}
\begin{equation}
\label{eq49}
v^\prime = b^{4 - d -\delta (1+\mu/2)} \left\{  v_0 -[4(n+8)v^2 -6v\mu]\ln b
 \right\} \;,
\end{equation}
\begin{equation}
\label{eq50}
\mu^\prime = b^{4-d} \left\{ \mu -[8(n+2)v\mu - 4\mu^2] \ln b \right\} \;.
\end{equation}
The extra--factor $b^{-\delta \mu/2}$ in Eq.~(\ref{eq49}) is not essential for
 the RG
analysis to this ($\tilde{\epsilon}^{1}$--) order of the theory because  $\mu
\delta \sim \epsilon \delta \sim  \epsilon^2$.

The RG relations~(\ref{eq48})--(\ref{eq50}) yield new $\delta$--corrections to
 the
relevant physical quantities.
They
 present
the difference between the critical properties of the impure film and the
corresponding infinite
impure system.

\subsection{FPs and critical exponents}
\label{sec6.2}

The RG relations~(\ref{eq48})~--~(\ref{eq50})
have four FPs:
the Gaussian FP (hereafter referred to as GFP) with coordinates
$(u_G=\mu_G=0)$, the so--called
"unphysical" FP (UFP), $v_{U} = 0$, $\mu_{U} = -\epsilon/4$,
 the Heisenberg FP (HFP),
\begin{equation}
\label{eq51}
v_H  = \frac{\epsilon - \delta}{4(n+8)} \;, \hspace{1.5cm} \mu_H =  0 \;,
\end{equation}
and the random FP (RFP),
\begin{equation}
\label{eq52}
v_R  =  \frac{\epsilon + 2\delta}{16(n-1)} \;,  \hspace{1.3cm}
\mu_R = \frac{(4-n)\epsilon + 2(n+2)\delta}{8(n-1)} \;.
\end{equation}
 All these FPs are known by the work of T. C. Lubensky~\cite{Lub:Rev}.
Here the FPs
require a further investigation because of the FS effect $(\delta > 0)$.

As usual, the GFP always exists and is stable for $d>4$. The UFP has no
physical meaning for
$\epsilon >0$, namely, for $d<4$, because the parameter $\mu$
must always be nonnegative ($\mu \sim \Delta$).
For $d>4$, the UFP is physical $(\mu_U > 0)$ but unstable. For
$4<d<(4+2\delta)$, the  UFP
has an instability towards $\mu$, whereas the same FP has a double instability
(towards both $v$ and $\mu$)
for $d>(4+2\delta)$.
Further, we shall concentrate our
attention on the GFP, HFP and RFP. In order to analyze their properties
 we must obtain the critical and stability exponents.

The stability exponents of the GFP are $\lambda_v^{(G)} = \epsilon - \delta$
and $\lambda_\mu^{(G)}=\epsilon$. These values show that the GFP is stable only
for $d>4$, where it describes an usual (free of disorder and fluctuation
interactions) Gaussian behaviour. The stability exponents of the HFP are
\begin{equation}
\label{eq53}
\lambda_v^{(H)} = \delta - \epsilon \;,
\end{equation}
and
\begin{equation}
\label{eq54}
\lambda_\mu^{(H)} = \frac{(4-n)\epsilon +2(n+2)\delta}{(n+8)} \;.
\end{equation}

The stability exponents $\lambda_{1,2}^{(R)}$ of the RFP are given by
\begin{equation}
\label{eq55}
\lambda_{1,2}^{(R)} = - \frac{1}{8(n-1)}
\left[ 3n\epsilon +2(4-n)\delta \mp \sqrt \Theta \right] \;,
\end{equation}
where
\begin{eqnarray}
\label{eq56}
\Theta & = &(5n-8)^2 \epsilon^2 -4(15n^2 +24n -48)\delta^2
\nonumber \\
& - & 12(n^2+12n-16)\epsilon \delta \;.
\end{eqnarray}
The static critical exponents describing the critical behaviour of
the  system are
given by $\eta=0$ and the value of the correlation length exponent $\nu$. For
the GFP
we have $\nu_G=1/2$, for HFP,
\begin{equation}
\label{eq57}
\nu_H = \frac{1}{2} + \frac{(n+2)}{4(n+8)} (\epsilon -\delta) \;,
\end{equation}
and for the RFP,
\begin{equation}
\label{eq58}
\nu_R = \frac{1}{2} + \frac{3n\epsilon + 2(n+2)\delta}{32(n-1)} \;.
\end{equation}
For $\delta=0$ and, hence, $D_{\mbox{\footnotesize eff}} = d$, the
Eqs.~(\ref{eq53})--(\ref{eq58}) yield the familiar results
for $d$--dimensional infinite impure systems~\cite{Lub:Rev}.
For the GFP and HFP we have
$z_d=1$, whereas for the UFP and RFP $z\neq1$. For the RFP Egs.~(\ref{eq47})
and~(\ref{eq52}) yield
\begin{equation}
\label{eq59}
z_d^{(R)}=1 + \frac{(4-n)\epsilon +2(n+2)\delta}{16(n-1)} \;.
\end{equation}

\subsection{Stability properties of the FPs}
\label{sec6.3}

Consider the stabilyty properties of the HFP.
The requirement of stability $\lambda_\mu^{(H)} < 0$ leads to the following
inequality
\begin{equation}
\label{eq60}
4(4-d+\delta) < (4-d-2\delta)n \;.
\end{equation}
When we solve this inequality with respect to $n$ we should have
in  mind that
the quantities $(4-d)=\epsilon \sim \delta$ are small and
that is why it is not convenient to divide the ineq.~(\ref{eq60}) by the small
factor  $(4-d-2\delta)$. For this reason, let us consider first the case
$\delta = 0$, where the inequality~(\ref{eq60}) is valid for $n>4$, provided
 $d<4$.
Further, we consider $n=4$ and for this value we easily find that the
 inequality~(\ref{eq60})
cannot be satisfied for any $\delta \geq 0$. Therefore, for $n=4$, the HFP is
unstable
towards the parameter $\mu$, i. e. towards the disorder effects. For infinite
systems ($\delta=0$) with $n=4$, the HFP has a marginal stability
($\lambda_{\mu}^{(H)} = 0$).
The problem of whether the HFP is stable for $n=4$ and $\delta=0$ has been
widely  discussed
in preceding studies~\cite{Lub:Rev} in high orders in $\epsilon$. In
particular, this topic has been extensively investigated by I. D.
Lawrie et al (see Ref.[12]) for the case of impure systems with a cubic
anisotropy. It has been shown that the higher orders in the
loop expansion do not give the initially expected reliability in the treatment
of the stability properties of the HFP near and at the value $n =4$ of the
symmetry index $n$. The reason is not in the specific features of the model but
is a general disadvantage of the RG studies of systems with competing
effects. Therefore, the same problem cannot be reliably
solved in the present case, too.

The inequality~(\ref{eq60}) is used together with $\lambda_v^{(H)} < 0$ and
$d>(2-\delta)$ in order to demonstrate that the HFP is unstable for all $n<4$.
For  $n > 4$,
the stability requirements $\lambda_v^{(H)} < 0$, $\lambda_\mu^{(H)} < 0$ and
$d>(2-\delta)$ yield the domain of stability
\begin{equation}
\label{eq61}
2-\delta < d <d_H \;,
\end{equation}
where
\begin{equation}
\label{eq62}
d_H = 4 - \frac{2(n+2)}{(n-4)} \delta \;.
\end{equation}
In fact, the conditions~(\ref{eq61}) are satisfied for all $\delta \in [0,1]$
 provided
$n > 16$, and for $0 \leq \delta < \delta_0$ with $\delta_0 = 2(n-4)/(n+8)$
 provided
$4<n<16$. Increasing the value of $\delta$, the interval of variations of $d$
 decreases and
for $\delta=\max (\delta_H,1)$ the width of the $d$--domain of stability becomes
 zero.

The stability properties of the RFP can be investigated only for $n \neq 1$,
because of
the degeneration~\cite{Lub:Rev} of the RG equations at $n=1$. This degeneration
leads to a
special critical behaviour described by
complex stability exponents~\cite{Lub:Rev} and, hance, by
oscillatory corrections to the main scaling laws~\cite{Lub:Rev}.
The investigation of the "oscillatory" critical behaviour is made in
 higher orders
of $\epsilon$ and $\delta$.
The values $0<n<1$ are of an academic interest only. We shall consider the case
 $n>1$.

Using the physical requirement $\mu_R > 0$, one obtains an inequality inverse
to~(\ref{eq60}). Further, we must distinguish between two cases
of stability of the RFP: (R) real negative exponents $\lambda_{1,2}^{(R)} < 0$,
and (C) complex exponets $\lambda_{1,2}^{(R)}$ with negative real parts.
 For the case (R), using
Eqs.~(\ref{eq55})-(\ref{eq56}), the requirement $\lambda_{1,2}^{(R)} <0$,
$d>(2-\delta)$ and the inequality inverse to~(\ref{eq60}), we determine the
following domain $R_R$ of stability in
the $(d,n,\delta)$ space:

\begin{equation}
\label{eq63}
2-\delta < d < (4+2\delta) \;,
\end{equation}

\begin{equation}
\label{eq64}
(4-n)(4-d) + 2(n+2) \delta > 0 \;,
\end{equation}

\begin{equation}
\label{eq65}
3n(4-d) + 2(4-n)\delta > 0 \;,
\end{equation}
and $\Theta \geq 0$. For the case (ii) of complex stability exponents
$\lambda_{1,2}^{(R)}$, the domain of stability $R_C$ is defined  by the
inequalities ~(\ref{eq64}), (\ref{eq65}), $\Theta < 0$, and $d>(2-\delta)$.

The criteria of stability can be investigated numerically for all allowed values
 of
$d, \delta$ and $n\neq1$. Here we shall restrict the analysis to those values of
 $n$ which
might be of importance for real systems or
for the explanation of the properties of the model.
The most important theoretical problem is
the comparison of the domains of stabilities of the G, H and R FPS for values
of  $n=2,3,4,..$.

For $n=4$, one easily obtains the picture in Fig. 3. The RFP with real
 exponents
$\lambda_{1,2}^{(R)}$ is stable in the domain $R_R$ defined by
 $0<\delta<\delta_4$
and $(2-d)<d<d_R(4)$, where $\delta_4=2/(1+2\sqrt 3)$ and $d_R(4) = 4 -
 2(1+\sqrt 3)$.
In Fig. 3, the point $a$ has coordinates ($\delta_a \equiv \delta_4
 \approx 0.45$,
$d_a = 1.55$), the point $b$ has coordinates $(1,1)$, and the point $c$ is
given  by $(1,4)$.
The domain $R_C$ in Fig. 3 denotes the stability region of the random
critical behaviour with complex stability exponents.
This domain is given by the inequalities $d_R(4)<d<4$ provided
$\delta<\delta_4$, and by $(2-\delta)<d<4$ for $\delta_4 < \delta \leq 1$.

For $1<n<4$ the stability criteria for $R$ in the
case of real stability exponents
are $\Theta \geq 0$ and
\begin{equation}
\label{eq66}
2-\delta < d < d_R \;,
\end{equation}
where
\begin{equation}
\label{eq67}
d_R = 4 + \frac{2(4-n)\delta}{3n} \;.
\end{equation}
The domain $R_{C}$ of complex stability exponents is given by the inequalities
$\Theta < 0$ and~(\ref{eq66}).

The domains of stability $R_R$ and $R_{C}$ for $1<n<4$ are quite similar to
 those in the
case $n=4$ depicted in Fig. 3, but the coordinates of the points $a$
 and
$c$ in the $(\delta,d)$ plane vary with $n$.
The coordinates ($\delta_a, d_a$) and ($d_c$, $\delta_c = 1$) of the points $a$
and $c$ are given
in Table 1 for several values of $n$.
The value $\delta_a$ decreases from $0.45$ to zero with the decrease of $n$ from
$n \sim 4$ to $n \sim 8/5$. For $n<8/5$, $\delta_a$ increases with the decrease
 of $n$.\\

TABLE 1. Values of $\delta_{a}, d_{a},$ and $d_{c}$\\
\begin{tabular}{|l|l|l|l|l|l|l|} \hline \hline
$n$ & 7/5 & 8/5 & 9/5 & 2 & 3 & 4 \\ \hline
$\delta_{a}$ & 1.06 & - & 0.02 & 0.05 & 0.26 & 0.45 \\ \hline
$d_{a}$ & 1.94 & - & 1.98 & 1.95 & 1.74 & 1.55 \\ \hline
$d_{c}$ & 5.24 & 5 & 4.81 & 4.67 & 4.22 & 4 \\ \hline  \hline
\end{tabular}\\

The value $n=8/5$ is quite special because for this value the function $\Theta
 (n,\epsilon,\delta)$
given by Eq.~(\ref{eq56}) depends linearly on $\epsilon$.
Within our one--loop approximation, the domain $R_R$ at $n=8/5$ does not exist
 at all
and the domain $R_C$ is expanded up to $0 \leq \delta \leq1$ and $2-\delta \leq
 d
\leq 4 + \delta$. In our case, the effective values of $\delta_a(8/5)$ and
$d_a(8/5)$ are $0$ and $2$, respectively.
It has been shown by I. D. Lawrie {\it et al}~\cite{Lub:Rev} that the same
value
$n=8/5$ is the reason for the peculiar critical behaviour of infinite impure
systems  $(\delta=0)$
with a cubic anisotropy. So,
 one may expect that in our
case a similar peculiar behaviour at
$n=8/5$ with singularities of the $\epsilon$--expansion and complex exponents
will occur.

It is seen from
Table 1 that the coordinate $d_c(n)$ in Fig. 3 is larger than
 $d_c(4)=4$
for all $0<n<4$. This means that for all $1<n<4$, there is a triangle domain
for certain values $d>4$,
where both the RFP and the GFP are stable. In this region there
is a competition between the pure
Gaussian behaviour represented by the GFP and the impure behaviour
represented by RFP.
The outcome of this competition should depend on the strength of the disorder
 effect.

For $n>4$, both the RFP and the HFP are stable for certain values of
 $d$ and
$\delta$. The stability domains $R_R$ and $R_C$ can
be
determined with the help of the inequalities~(\ref{eq63})--(\ref{eq65}).
For $\delta < \delta_0$, the domain
$R_R$ is confined by the inequalities $\Theta \geq 0$, and
\begin{equation}
\label{eq68}
d_H < d < d_R
\end{equation}
whereas for $\delta>\delta_0$, the same domain is described by the inequalities
$\Theta \geq 0$ and
\begin{equation}
\label{eq69}
2-\delta < d < d_R
\end{equation}
 The domain $R_{C}$ is defined by the same inequalities
(\ref{eq68}) and~(\ref{eq69}) but for $\Theta < 0$.

A typical picture of the stability domains
in this case is given in Fig. 4 for $n=6$, where the points $h$ and $a$
 have
coordinates: $\delta_h \equiv \delta_0 (6) = 0.29$, $d_h = 1.61$, $\delta_a =
 0.74$ and
$d_a = 1.26$. The HFP is stable for relatively small values of $\delta$,
whereas the  stability
of the RFP dominates for large values of $\delta$. There is a domain
$(4-0.22\delta) < d < 4$ of dimensionalities $d \sim 4$ for which both the RFP
and the HFP are unstable. This domain is defined by
the  location of the
point $c$ with coordinates $\delta_c = 1, d_c(6)=3.78$; see the shaded region
in Fig. 4.
Such domains of instability exist for all symmetry indices $n>4$ and $\delta >
 0$.
They can be described by the inequality $d^\prime (n,\delta) < d< 4$ which can
 be
obtained by the RG analysis. For $n=6$ and $\delta = 1$, $d^\prime (n,\delta) =
 d_c(6)$
as shown in Fig. 4. For $\delta \to 0$, $d^\prime (n,\delta) \to 4$ and
 for
$\delta \to 1$, $d^\prime (n,\delta)$ is lowered up to a minimal value
$d^\prime (\infty,1)=3.33$ (See also Fig. 5).

As the GFP is stable only for $d>4$, the shaded domain in Fig. 5
remains the region of a total instability of the system. This total lack of
stable FPs for all $n>4$, $\delta > 0$ and certain $d \sim 4$
can be interpreted as an indication for a fluctuation--driven impure tricritical
phenomenon followed by a first--order phase transition. The FS system ($\delta
> 0)$ is unstable towards the disorder (the parameter $\mu$) and, hence, the
reason for the
appearance of a fluctuation-driven phase transition of first--order is
in the simultaneous effect of the FS and the extended impurities.

In order to justify the
prediction for the existence of a tricritical point and a fluctuation (or
 disorder)--driven
phase transition of first order, we remind that the fluctuation interactions
 represented
by the parameter $u$ are relatively small near four dimensionalities
 $(d\sim4)$.
Under such circumstances, the disorder
 effects may
alter the sign of the effectivelly small parameter $u$ from $u>0$
(a second--order transition)
to $u<0$ (a first--order transition). At $u=0$, a tricritical point should
appear.  Note, that these
effects are impossible for infinite impure systems and in pure FS systems.
 The change of the order
 of the phase
transition in the present case is a result of the competition between
 fluctuation, disorder
and FS effects. The effective fluctuation model for the description of this
change of the order of
the phase transition can be constructed by adding an
additional $u_6 \phi^6$ term to the Hamiltonian~(\ref{eq1}).

The stability domain of the HFP vanishes when the symmetry index $n$ decreases
from $6$ to $4$.
Increasing the number $n$ to $n=16$, the stability region of the HFP expands up
to the triangle $2b4$ (see the dotted line $4b$ in Fig. 4).
The further increase of the values of $n$ above $n=16$ leads to an enlargement
of the stability region of the HFP so that the points $h$ and $a$ already lay
on the  vertical
line $\delta=1$.  For $n \to \infty$ (the limit of the spherical
 model~\cite{Uzun:Int}) the point
$h$ has coordinates $(1,2)$ on the vertical line $\delta=1$, as shown in
 Fig. 5.
The $R_R$ is extremely small but persists (see the narrow wing $h4a$ with $d_a =
 2.19$).
The domain $R_C$ is shown in Fig. 5 by the triangle $a4c$, where $d_c =
 3.33$.
In Figs. 4--6, the point $b$ has the same coordinates $(1,1)$.

\section{Summary and validity of the results}
\label{sec7}

We have obtained and analyzed in details the RG recursion relations for FS
 systems
of slab geometry described by the $n$--component LG model containing quenched
impurities. Two models of quenched impurities have been considered.
The results demonstrate the outcome of the competition between
the effects of fluctuation interactions, the finite size, and the disorder.

We have used an $\tilde{\epsilon} = (4 - D_{\mbox{\footnotesize
eff}})$--expansion defined by the effective spatial dimensionality $\delta =
(D_{\mbox{\footnotesize eff}} - d)$ of the
fluctuation modes of the order parameter. Our approach is based on an
approximate substitution of the lattice summation with an integration in
dimensionalities less than unity. The limitations of this approach have been
considered. The approach is more general that the familiar RG
considerations based on the $\epsilon = (4-d)$--expansion. The fundamental
relationship between
the effective dimensionality $D_{\mbox{\footnotesize eff}}$ and the ratio
$y = (L_{0}/\xi)$ of the characteristic lengths of the film has been introduced
and discussed.

The most interesting cases of critical phenomena are those for $\delta \ll 1$
(thin films)
and $\delta \gg 1$ (thick films, when the film behaves as an almost--infinite
system). It has been demonstrated that our approach based on an integration in
dimensionalities less than unity is reliable for the description of these
cases, in particular, for real films $(D = 3)$.

The method introduced in this paper has been used to demostrate the
validity of the FSC in films with homogeneously distributed quenched impurities
with short--range random correlations. It has been shown that the proof of the
FSC in these systems can be easily proven and straightforwardly expanded to any
order in the loop
expansion. The reason is that the impurity correlations are spatially isotropic
and the sum ($\epsilon + \delta$) can be taken as the single expansion
parameter $\tilde{\epsilon}$.

The extended impurities break the spatial isotropy of the disordered system,
and the RG investigation yields a dependence of the critical exponents on two
small parameters: $\epsilon = (4-d) = (5 - D)$ and $\delta$. This leads to
another picture of the critical behaviour which will be summarized in the
remainder of this Section.

The dependence of the
FP coordinates and the critical exponents on the small parameters ($\epsilon,
\delta$) and the symmetry
index $n$ has been calculated and analyzed with respect to the stability
properties of the FPs. In this way we have established the possible types of
critical behaviour for different values of $\delta$, the geometric
dimensionality $D$ of the hyperslab, and $n$. In order to describe
the random
critical behaviour in films with extended impuirities we have introduce and
calculated a new critical exponent $z_{d}$. The analogy between this exponent
and the dynamical exponent in quantum systems has been emphasized. The results
are valid for effective dimensionalities
$D_{\mbox{\footnotesize eff}} > D^{(L)}_{\mbox{\footnotesize eff}} =  2$ -- the
lower borderline dimensionality.

For real films $(D = 3)$ with extended impurities our
analysis is valid for the so--called quasi--$2D$ (thin films), where $\delta >
0$. So, the results from our RG analysis performed for small values of
$\epsilon$ and $\delta$ can be exptrapolated for predictions of the critical
behaviour of real impure films ($D=3, d=2$) with a finine thickness $L_{0} \gg
a$, where the
$\delta(y) > 0$. We have shown that, in the strict mathematical sense, the
latter condition is always valid for
$y > 0$, i. e. when the system is outside the critical point. But in real
expriments the conditions for the validity of our consideration are somewhat
different. The function $\delta(y) \sim 0$ up to values of $y \sim 1$ and,
therefore,
in experimental conditions ($\xi \gg a$) the thickness $L_{0} \sim \xi$ should
be much larger than the interparticle distance $a$.  As far as the
behaviour in the critical region ($\xi \gg a)$ is concerned the condition $y >
0$ will be easily satisfied for films with $L_{0} \gg a$. This is the
case of interest for experiments. The case of exactly $2D$ films $(L_{0} = a)$
corresponds to the lower borderline dimensionality and, therefore, is beyond
the scope of our investigation.
The invalidity of the results at the lower borderline dimensionality is a
generic disadvantage
of the field--theoretical RG methods rather than a result of our particular
approach to the problem.
The $2D$ case can be effectively achieved for any film in the
asymptotic vicinity of the critical point, where $\xi \to \infty$ and, hence,
the ratio $y \to 0$ for any finite thickness $L_{0}$. However, this situation
is not of interest for the most part of real experiments.

While the condition $D_{\mbox{\footnotesize eff}} >
D^{(L)}_{\mbox{\footnotesize eff}} = 2$ guarantees the reliability of the
qualitative predictions of our RG analysis, the extrapolation
of the results obtained near the upper borderline dimensionality
$D^{(U)}_{\mbox{\footnotesize eff}} = (4 - d -\delta)$ to the real
dimensionality $(d = 2)$ requires an extension of the values of the small
parameter $\tilde{\epsilon}$ to values $\tilde{\epsilon} \sim{(2-\delta)} < 2$.
This is the
usual way of extrapolation of the RG results to real systems. It is believed
the small parameter $\tilde{\epsilon}$ should be set equal to ($2 - \delta$)
in order to predict the values of the critical exponents for the real case
($\delta \ll 1$) and that in this case, the one--loop results do not give
quantitatively correct
predictions. Although this is generally true, the qualitative predictions from
the one--loop results are reliable, in particular, for the solution of
important problems
like the possible types of critical behaviour and the conditions under which a
critical behaviour can occur. In this article we have been mainly involved in
these type of
problems and the results are well grounded within the one--loop approximation.
The practice of the numerous applications of the RG to complex systems shows
that the one--loop results about the general picture of the critical phenomena
remain valid within the framework of consideration by higher loop
approximations. For real films $\tilde{\epsilon} =
(2 - \delta)$ should take values $\tilde{\epsilon} \sim 2$ for $\delta \ll 1$
and $\tilde{\epsilon} \sim 1$ for $\delta \sim 1$ (thick films). Therefore, the
analysis carried out in Section 6 is reliable for the case of real films and
their critical behaviour is given by Figs. 3~--~5 on the
line $d = 2$.
In addition
to this conventional point of view we shall stress that the results for small
values of $\epsilon = (4 - d)$ can also be extrapolated to real films but for
those parts of the critical region which are not asymptotically close to the
critical point. Thus the fluctuation--driven change of the order of the phase
transition near the upper borderline dimensionality $(4 + \delta)$ can be
extrapolated to real effective dimensionality $(d + \delta)$, provided one is
interested in fluctuation phenomena which occur outside the asymptotic vicinity
of the critical point.

A brief summary of other results for the critical behaviour
in real dimensionalities $(d = 2, 0 <
\delta < 1)$ can be outlined with the help of Figs. 3~--~5. This critical
behaviour is described by the HFP and the RFP.
For $1 < n \leq 4$, only the random critical behaviour is present. For
relatively
thick films this critical behaviour will exhibit oscilatory corrections to the
scaling laws described by complex stability exponents, whereas the ramiliar
random critical behaviour will occur in relatively thin films. For $n > 4$, the
pure critical behaviour represented by the properties of the HFP will occur for
relativelu thin films whereas the random critical behaviour in its two variants
of real and complex stability exponents will occur in thick films. In the
spherical limit $( n = \infty)$ the pure critical behaviour is stable for all
real films and the disorder effects are irrelevant for the critical properties.
The singular behaviour of the films for the symmetry indices $n = 1$ and $n =
8/5$  has been explained in Section 6.

Finally, we note,  that
the present investigation reveals essential differences in the critical
behaviour of infinite
and FS systems with extended impurities. The finite
 thickness $L_0$
of the film leads to: an essential dependence of the critical exponents on the
 noninteger
dimensional portion $\delta$, a fluctuation--driven change of the order of the
 phase
transition for all $n>4$ and $d \sim 4$ (the almost--MF region), the
competition between the  pure
(Gaussian) and impure critical regimes for all $1<n<4$ and $2 \leq d < 4$, and
the dominating role
of the impure (random) critical behaviour with oscillatory
 corrections to
the scaling laws for all relatively thick films.

{\bf Acknowledgments:}

L.~Craco and D.~I.~Uzunov gratefully acknowledge the hospitality of the
MPI-PKS (Dresden),
Aussenstelle Stuttgart and, in particular, the kind support by Prof. Peter
 Fulde.
L. Craco thanks the support by the Brazilian Agency Conselho Nacional de
Desenvolvimento Cient\'{\i}fico e Tecnol\'ogico (CNPq). A research
grant Ph560 of NFS (Sofia) is also acknowledged.

\newpage

{\bf Figure captions}

Fig.1.  First--order diagram for $k'_{0}$. The discontinuous line reprsents the
disorder average\\ $[|\varphi(\vec{k})|^{2}]_{R} = \Delta$ and the continuous
line represents the bare correlation function\\ $G_{0}(k_{0},\vec{k} +
\vec{k}_{1})$.

Fig.2. Stability domains of $R$ for $n = 4$: $R_{R}$ (triangle $2a4$) and
$R_{C}$ ($4abc$).

Fig. 3. Stability domains of $H$ (triangle $2h4$) and $R$ for $n > 4$. $R_{R}$
is given by the triangle $h4a$ and $R_{C}$ is the rectangle $a4cb$. The dotted
line $b4$ marks the extension of the stability domain of $H$ for $n = 16$. The
shaded region is explained in the text.

Fig. 4. Stability domains of $H$ and $R$ for $n = \infty$ (Hartree limit [10]).
$R_{R}$ and $R_{C}$ are given by the triangles $h4a$ and $a4c$, respectively.

 \end{document}